# Probing the degree of spin polarization of a ferromagnet with ferromagnet/superconductor proximity effect

Pavel V. Leksin[1,2], Andrey A. Kamashev[2], Joachim Schumann[1], Vladislav Kataev[1], Jürgen Thomas[1], Thomas Gemming[1], Bernd Büchner[1,3], and Ilgiz A. Garifullin[2]

[1]*Leibniz Institute for Solid State and Materials Research IFW Dresden, D-01069 Dresden, Germany*
[2]*Zavoisky Physical-Technical Institute, Russian Academy of Sciences, 420029 Kazan, Russia*
[3]*Technical University Dresden, D-01062 Dresden, Germany*

**Abstract**

Superconductor/ferromagnet proximity effect has been studied for Pb/Co$_2$Cr$_{1-x}$Fe$_x$Al bilayers. Different substrate temperatures allowed us to prepare the Heusler alloy Co$_2$Cr$_{1-x}$Fe$_x$Al films with different degree of spin polarization of conduction band of the Heusler layer. We obtain a strong correlation between the dependence of the superconducting transition temperature on the Pb-layer thickness at fixed thickness of the Heusler layer and the degree of spin polarization in the ferromagnetic layer.

The effectiveness of magnetoelectronic devices depends on the extent to which a current is spin-polarized [1]. A typical transition-metal ferromagnet has two components to its electronic structure: narrow *d* band that may be fully or partially spin polarized and broad *s* band with a lesser degree of spin polarization (DSP) (due to hybridization with the *d* band). The knowledge of DSP gives an opportunity to distinguish between an ordinary ferromagnet and a half-metal (ferromagnet with 100 % DSP). Half-metallic ferromagnets represent the class of materials which have recently attracted a considerable interest due to their possible applications in spin electronics. In these materials the two spin bands show a completely different behavior. Usually the majority band (spin-up) shows a typical metallic behavior with a nonzero density of states (DOS) at the Fermi level $E_F$ and the minority (spin-down) band exhibits a semiconducting behavior with a gap at $E_F$. Therefore, such half-metals (HM) can be considered as hybrides between metals and semiconductors.

Some Heusler alloys of Co$_2$YZ composition (where Y is a 3*d* transition metal and Z is an *s-p* metal) are ferromagnets and are expected to exhibit 100 % spin polarization due to the energy-band gap for the minority–spin electrons at the Fermi level ($E_F$) (see, e.g., [2]). DSP of these alloys and their films strongly depends on the preparation conditions [3, 4]. In order to get a ferromagnetic film with DSP approaching the meaning of half-metal usually it is necessary to hold substrate during the film growth at temperatures well above the room temperature.

For both scientific and technological reasons it is important to be able to directly and easily measure DSP of a ferromagnet. The most natural and common definition of DSP of a conduction band of ferromagnet is [5]:

$$P = \frac{N_\uparrow(E_F) - N_\downarrow(E_F)}{N_\uparrow(E_F) + N_\downarrow(E_F)}, \qquad (1)$$

where $N_{\uparrow(\downarrow)}(E_F)$ is the density of electronic states at the Fermi level with corresponding spin direction ($\uparrow, \downarrow$).

The following direct methods are usually involved for investigating DSP. Typical experiments that can probe $P$ are spin resolved photoemission spectroscopy [6] and hard X-ray photoemission spectroscopy [7, 8]. However, their resolution is hundreds of meV, which is significantly less than the necessary one (~1 meV). They also imply complicated apparatus such as synchrotron radiation, and very stringent surface preparation. A very useful method is the tunneling spectroscopy. The pioneering experiments by Tedrow and Meservey [5, 9] have shown, that the tunnel junction for ferromagnet/ insulator/superconductor (F/I/S) can be used to measure the DSP. One can also measure tunneling currents separately for both spin polarization channels with superconducting point contacts. However, to get precise results, the barrier and interface quality requirements are severe, so the measured DSP correlates with the junction structural quality. Another technique, based on the combination of tunneling magnetoresistance (TMR) and F/I/F junction-Julliere's model, is widely used for estimating the DSP [10]. It is clear, however, that the results depend on the structural quality of the tunnel junction and the choice of the tunneling barrier. It was also suggested [11, 12] that Andreev's reflection at the interface between ferromagnet and a superconductor can be used for direct probing of DSP. The advantage of such relatively novel device-independent technique is its experimental simplicity. Unlike other methods, which have stringent requirement for the surface atomic cleanness and/or an ultrathin uniform oxide layer, and thus may make the study of some interesting materials difficult, the superconducting point contact method requires no magnetic field and has no special constraints on a sample. The S/F point contact is usually organized between the surface of the sample and a superconducting probing element, a sharp needle, for instance. At small voltages the differential conductance of such a contact decreases with the rise of the DSP value, which makes it advantageous for a routine optimization of DSP for the material. The results can be modeled by using different models, such as the Blonder-Tinkham-Klapwijk model [13] and its modifications (see, eg., [14-16]). However the theoretical calculations sometimes fit results unsatisfactorily [17] or produce unrealistic fitting parameters. The data which come from all the described methods were analyzed by Mazin [18]. Most of the above described methods of measuring DSP require clean and well-ordered surfaces, which for Heusler compounds is difficult to obtain by surface cleaning procedures of *ex situ* prepared samples.

In this paper we show that the dependence of the superconducting transition temperature on the superconducting layer thickness in a superconductor/ferromagnet (S/F) bilayer dramatically changes with changing DSP. This suggests a simple alternative method for estimation of DSP of the ferromagnetic layer. In contrast to other techniques such method does not involve the top clean surface or junction effects and thus may provide the information about DSP in simple bilayer geometry. In S/F bilayers the singlet Cooper pair wave function penetrates from superconductor to ferromagnet over a certain distance, which is usually associated with the penetration depth for ferromagnet depending on the exchange splitting of conduction band of ferromagnet. This process is usually accompanied by a pair breaking effect, which decreases $T_c$ of the S layer and leads to a complete suppression of the superconductivity at a certain critical superconductor layer thickness $d_S^{crit}$.

We show the efficiency of our method using the Heusler alloy with the nominal composition of the target $Co_2Cr_{0.4}Fe_{0.4}Al_{1.2}$. This choice is determined by the possibility to change easily DSP of conduction band by changing the substrate temperature during the growth of the Heusler alloy layer [3, 4].

It is known [3] that the alloy $Co_2Cr_{1-x}Fe_xAl$ forms the film with high DSP if during the film growth the substrate temperature is hold at $T_{sub} \geq 600$ K. The films prepared at lower $T_{sub}$ appear to be weak ferromagnets due to disordered structure [3]. According to our data on the point-conctact spectroscopy for the studied samples, DSP reaches 70 %. Bearing this in mind we prepared two sets of MgO/Heusler(12nm)/Cu(1.5nm)/Pb($d_{Pb}$) samples with variable Pb-layer thickness and with the substrate temperature $T_{sub}$ = 300 K (Set 1) and 600 K (Set 2) when evaporating the Heusler alloy by sputtering technique. Here we used Cu(1.5nm) as antidiffusion layer. To optimize the growth of the Cu/Pb fragment after deposition of the Heusler layer we decreased the temperature of the substrate $T_{sub}$ down to 150 K. An advantage of the low $T_{sub}$ was shown in our previous papers [19-20]. Finally all samples were covered by $Si_3N_4$ protective layer against oxidation. We used the following deposition rates: 0.37 Å/s for Heuser alloy, 0.5 Å/s for Cu layers and 12 Å/s for Pb films.

The deposition of layers was performed using a combination of the sputtering technique

(for Heusler alloy) and an e-gun in ultra-high vacuum (UHV) with pressure $10^{-9}$ mbar (for Cu and Pb). The deposition setup had a load lock station with vacuum shutters, allowing changing the sample holder without breaking the UHV in the main deposition chamber. First, the substrates were fixed on a sample holder and transferred into the sputtering chamber for deposition of the Heusler alloy using the dc sputtering technique. Then the sample holder was moved to the main deposition chamber through the load lock station. We used a rotating wheel sample holder in order to prepare a set of samples with different Pb-layer thickness in a single vacuum cycle.

To inspect the layer stacks regarding the thickness of the layers as well as the interface roughness and the morphology of the Pb layer, cross sections of the samples were investigated with a transmission electron microscope FEI TEM/STEM Tecnai F30 working at an acceleration voltage of 300 kV. The electron-transparent lamellas were prepared by the focused ion beam

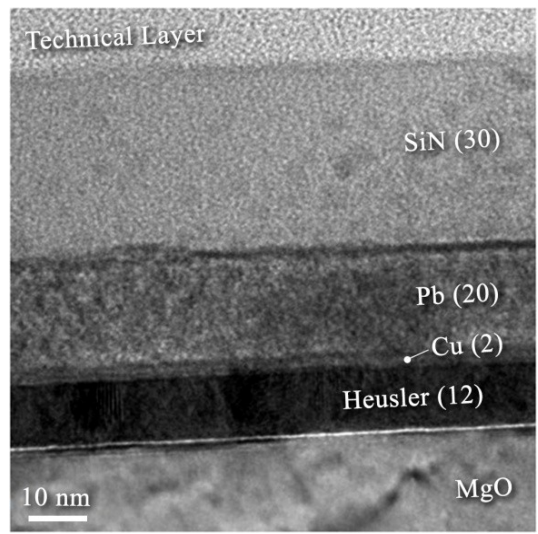

**Figure 1** Microscopic cross-sectional image of the Heusler/Cu/Pb sample with $T_{sub}$ = 300 K. This image is typical also for Heusler(12)/Cu(1.5)/Pb(20) with $T_{sub}$ = 600 K.

(FIB) technique using a Zeiss 1540XB cross beam machine. After inspection of imaging a C/Pt-O-H protection layer (technical layer in Fig. 1) was deposited at the position of interest. The lamella was cut by a focused 30 keV Ga ion beam and after its lifting out welded on an electron microscopic girder. The protection layer reduces the Ga implantation in the sample region close to the surface. The cross sections were analyzed in the TEM by means of conventional fixed beam imaging including its high resolution option as well as in the scanning transmission electron microscopic mode (STEM) using a high angle annular dark field detector (HAADF). Additionally, by energy dispersive X-ray spectroscopy (EDXS) in analytical mode the composition $Co_{61}Cr_{13}Fe_{11}Al_{15}$ for the samples prepared at $T_{sub}$ = 300 K and $Co_{55}Cr_{15}Fe_{13}Al_{17}$ at $T_{sub}$ =

600 K was determined. The composition in both cases is nearly the same with some deficiency of Al. The interfaces between the single layers could clearly be seen in the TEM micrographs as well as in the STEM-HAADF images (Fig. 1). In both structures prepared at $T_{sub}$ = 300 K and 600 K all thicknesses are nearly the same and the interfaces are flat. The EDXS linescan intensity profiles for the cross-sections revealed no chemical differences of the interfaces Heusler/Cu and Cu/Pb for both types of Heusler films, evidencing the absence of the oxide layers in both cases. This fact was checked especially carefully.

Both structures were magnetically characterized using a standard 7 T VSM SQUID magnetometer. Magnetic hysteresis loops were measured at $T$ = 10 K with magnetic film in the film plane (see Fig 2). One can see that the saturation magnetization of 850 emu/cm³ for the sample from Set 2 prepared at $T_{sub}$ = 600 K is larger than that for sample from Set 1 (570 emu/cm³). Note, that for pure iron this value is equal to 1730 emu/cm³. According to the data by Miura et al. [21] Co ↔ Cr disorder (i. e. the appearance of Cr atoms at the Co sites and vice versa) significantly reduces the total magnetic moment and DSP. Thus, our data are in a good agreement with the data by Miura et al.

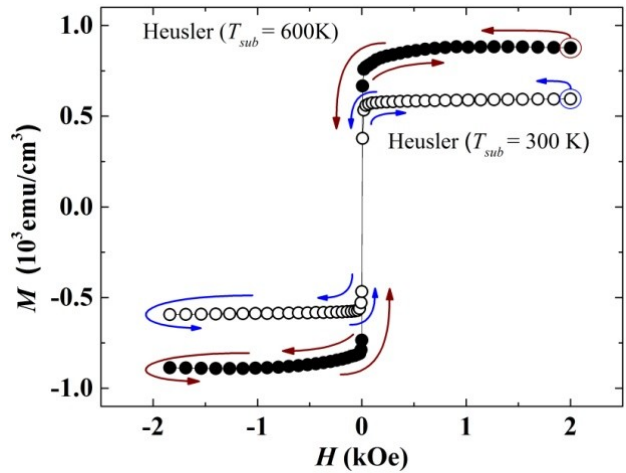

**Figure 2** Magnetic hysteresis curves for Heusler(12) with $T_{sub}$ = 300 K (open circles) and Heusler(12) with $T_{sub}$ = 600 K (closed circles). The measurements were carried out at $T$ = 10K. Arrows indicate the direction of the magnetic field sweeps.

The transport properties of the samples were studied using a 4-contact resistivity measurement method using the B2902A precision source/measure unit from Keysight Techn. The temperature of the sample was controlled with the 230 Ω Allen-Bradley thermometer which is particular sensitive in the temperature range of interest. We found that the residual resistivity ratio $RRR = \rho(300\ K)/\rho(10\ K)$ of the studied samples lies in the interval 10 < $RRR$ < 17. Using $\rho(300\ K)$ = 21 μΩ·cm [22] we obtain $\rho_s = \rho(10\ K)$ = 1.2 - 2.1 μΩ·cm for the residual resistivity. The BCS coherence length for Pb amounts to $\xi_0$ = 83 nm [22] and the mean-free path of conduction electrons obtained using the Pippard's relations [23] is about $l_s \sim 17$ nm. The comparison of $l_s$ with $\xi_0$ shows that $l_s \ll \xi_0$ implying the "dirty" limit for the superconducting part of the system. Therefore, we calculate the superconducting coherence length as $\xi_s = \sqrt{\xi_0 l_s/3.4} = 41$ nm. We also performed the measurements of the temperature dependence of the resistivity of the single films of the Heusler alloy prepared at $T_{sub} = 300$ K and 600 K. For the films prepared at $T_{sub} = 300$ K we obtain that the resistivity does not depend on temperature and amounts to $\rho_f = 143$ μΩ·cm (cf. 220 μΩ·cm in Ref. [3] and 170 μΩ·cm in Ref. [4]). For the film prepared at $T_{sub} = 600$ K this value is also independent of temperature and amounts to $\rho_f = 130$ μΩ·cm (cf. 330 μΩ·cm in Ref. [3] and 170 μΩ·cm in Ref. [4]).

We have measured $T_c(d_{Pb})$ for all prepared samples. The obtained results are shown in Fig. 3(a). The shape of the dependence of $T_c$ on the Pb-layer thickness $d_{Pb}$ at fixed thickness of the Heusler layer thickness is conventional: with decreasing Pb thickness, $T_c$ decreases slowly at large $d_{Pb}$ and then drops sharply to zero when $d_{Pb}$ approaches the critical thickness $d_{Pb}^{crit}$. Our previous results for Fe/Cu/Pb structure [24] are also shown for comparison. One can see that the largest $T_c$-suppression as function of $d_{Pb}$ is observed for Fe/Cu/Pb structures. For Set 1 prepared at $T_{sub} = 300$ K the suppression is weaker [closed circles in Fig. 3(a)].

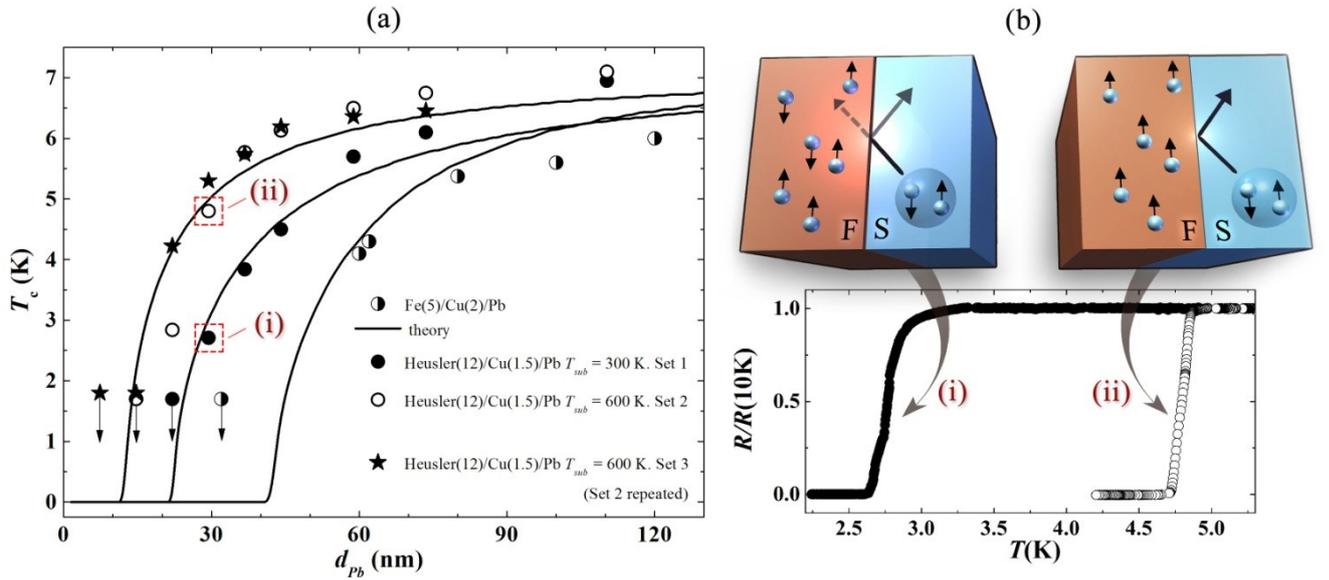

**Figure 3** (a) The $T_c$ dependence on the superconducting Pb layer thickness $d_{Pb}$ for: Fe(5nm)/Cu(2nm)/Pb($d_{Pb}$) [24] (half-opened circles); Heusler(12nm)/Cu(1.5nm)/Pb($d_{Pb}$) with $T_{sub} = 300$ K (closed circles); two independent sets of samples Heusler(12nm)/Cu(1.5nm)/ Pb($d_{Pb}$) with $T_{sub} = 600$ K (opened circles and black stars). (b) Superconducting transition curves for samples Heusler(12nm)/Cu(1.5nm)/ Pb(30) with $T_{sub} = 300$ K (i) and $T_{sub} = 600$ K (ii). The sketch visualizes the process at the F/S interface for samples with small (left) and large (right) DSP, respectively.

Finally, for Set 2 prepared at $T_{sub}$ = 600 K the $T_c(d_{Pb})$-curve [open circles in Fig. 3(a)] is shifted further to the left from the data for Set 1. To confirm the latter result we prepared the control set of samples at $T_{sub}$ = 600 K [Set 3 black stars on Fig.3 (a)]. The obtained results coincide perfectly with the data for Set 2. Fig. 3(b) demonstrates the shift of the superconducting transitions curves for two samples from Set 1 (i) and Set 2 (ii). Both have the same layer stack composition, but the $T_c$ values differ dramatically [cf. Fig. 3 (a)]. The $T_c$ difference between (i) and (ii) is about 2.1 K. When increasing $T_{sub}$, the shift of $T_c(d_{Pb})$-curve occurs not due to the change of the saturation magnetization (Fig. 2). Since $M_{Set2}^{sat} > M_{Set1}^{sat}$ (Fig. 2), the $T_c(d_{Pb})$-curve for Set 2 should be shifted then to higher thicknesses $d_{Pb}$, as it indeed the case for the Fe/Cu/Pb structure [Fig. 3(a)], which has the largest saturation magnetization among the studied samples. Furthermore, the composition of the Heusler layer and the chemical composition of the Heusler/Cu and Cu/Pb interfaces are the same for both types of the Heusler-based structures and thus could not be the reason for the significant shift of the $T_c(d_{Pb})$ dependence.

Therefore, we argue that a substantial offset of the $T_c(d_{Pb})$-curves may be caused by different DSP of the Heusler part in samples Set 1 and Set 2. The larger DSP of the F - layer in the S/F - heterostructure should inhibit the penetration of Cooper electron pairs with opposite spins from the S - layer into the F - layer because of the different density of states of the up-spin and down-spin electrons in the F - layer. Thus, the suppression of $T_c$ should take place at smaller thicknesses of the S - layer. To verify experimentally this conjecture we have studied our Heusler films, prepared at $T_{sub}$ = 300 K and 600 K, with Andreev's reflection point contact spectroscopy. Similar to Soulen *et al.* [12] the tunnel junction technique has been successfully used to compare the spin polarization DSP for the studied Heusler films. The general drawback of this technique is the constraint on the fabrication of a device consisting of the oxidized Heusler film. We have formed a metallic point contact between the sample and a superconducting needle using a simple mechanical adjustment. The needle was driven by a micrometer piezo mechanism. A metallic contact allows coherent two-particle transfer at the interface between the normal metal and a superconductor. The electronic transport properties at the point contact measures the conversion between superconducting pairs and the single-particle charge carriers of the metal. We used Nb tips as the probing needles. Special care was taken to prevent Nb oxidation during the needle preparation in air. The transport measurements were performed using the standard four-contact technique while the point contact and sample were immersed in liquid helium. The $dI/dU$ data in this study was obtained by a standard *dc* technique. The resistance of the measured Nb/Heusler contacts was around 60

Ohms for both types of the samples. The contacts were tested first at temperature $T = 15$ K which is higher, than the $T_c = 9.25$ K for Nb and have shown a constant differential resistance at any $U$ value. The results measured at $T = 4.2$ K are plotted in Fig. 4 (a). The conductance $dI/dU$ of sample from Set 1 ($T_{sub} = 300$ K) is maximum at the voltage $U \approx \pm 1$ V and continuously decreases with either increasing or decreasing $U$. In contrast, $dI/dU$ of sample from Set 2 ($T_{sub} = 600$ K) is practically constant at $|U| \geq 3.5$ V and exhibits a minimum at $U = 0$.

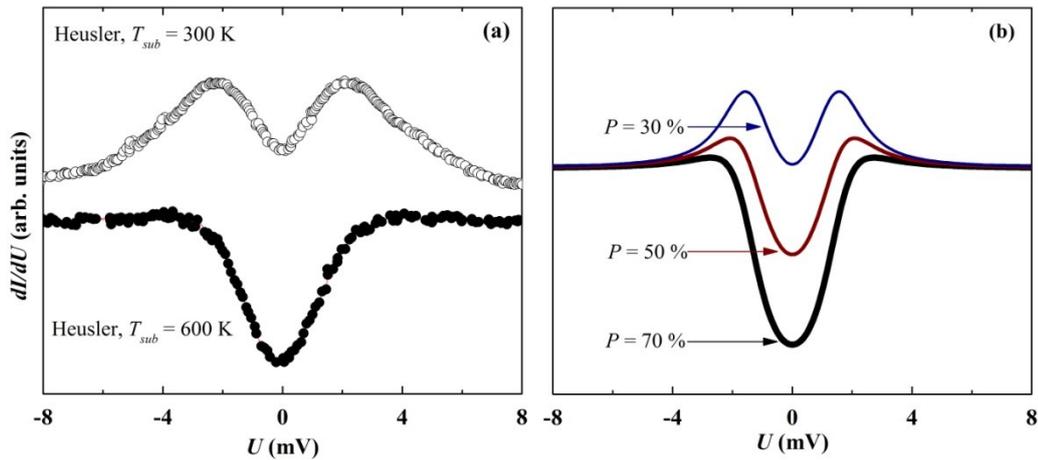

**Figure 4** (a) Andreev's reflection point contact spectra for: (a) Heusler with $T_{sub} = 300$ K (open circles) and Heusler with $T_{sub} = 600$ K (closed circles). The thicknesses of the films were 200 nm; (b) theory by G. I. Strijkers et at. [14].

If $P = 100$ % near $E_F$, then there are no spin-down states in the Heusler-film to provide the spin-down electron of the superconducting pair for Andreev reflection. Supercurrent conversion via Andreev reflection at the interface is effectively blocked, allowing only single-particle excitation to contribute to the conductance. These single-particle states necessarily see the gap in the energy spectrum of the superconductor, thus suppressing the conductance for energies less than exchange splitting of the conduction band of the Heusler-film.

The calculated $dI/dU$ dependences for different $P$ values of DSP using a classical theoretical model by Strijkers et al. [14] are plotted in Fig.4 (b). As can be seen from Fig. 4 (b), the curves at small voltages exhibit the tendency to rise up as the P rises from 30 to 70 %. Here we set the temperature for the system $T = 4.2$ K, S/F interface transparency parameter $Z = 0.4$ and the superconducting energy gap $\Delta = 1.5$ meV. The experimental curves on Fig. 4 (a) exhibit exactly the same tendency as the calculated ones, evidencing that the DSP value for the sample from Set 2 is higher, than the one for the sample from Set 1. First, according to theory [see Fig.4 (b)] the $dI/dU$ for DSP = 30 % monotonically rises up as the voltage amplitude is varied from 8 mV to 2 mV. In contrast, the $dI/dU$ curve for DSP = 70 % is almost constant within this voltage range.

Second, the curve for DSP= 30 % at $|U| < 1.5$ mV has two maxima and one minimum. At high polarization values about DSP = 70 % the curve has one pronounced minimum with the strongly damped maxima. The experimental curves in Fig. 4 (a) show qualitatively the same characteristic changes, suggesting the larger value of DSP ~ 70 % for Heusler film, prepared at $T_{sub}$ = 600 K, as compared with DSP ~ 30 % of the Heusler film prepared at $T_{sub}$ = 300 K.

With this knowledge let us discuss our results on the $T_c(d_{Pb})$ dependences (Fig. 3). In accordance with the theory by Fominov et al. [25], the explicit result for the critical thickness of the S - layer $d_S^{crit}$ can be obtained in the limit $(\gamma/\gamma_b)(d_s/\xi_s) \ll 1$ as

$$\frac{d_S^{crit}}{\xi_S} = 2\gamma_E \left(\frac{\gamma}{\gamma_b}\right), (2); \quad \gamma = \frac{\rho_S \xi_S}{\rho_F \xi_F}, (3); \quad \gamma = \frac{R_b A}{\rho_F \xi_F}. (4)$$

Here $\gamma_E \approx 1.78$ is the Euler constant, $\rho_S$ and $\rho_F$ are the normal-state resistivity of the S - and F - layer; $R_b$ is the normal-state resistance of the S/F boundary and $A$ is its area. The value of $\gamma_b = 0$ corresponds to the fully transparent S/F interface. $d_S^{crit}$ is smaller for larger values of $\gamma_b$.

The $T_c$ curve in Fig. 3(a) show that the critical thickness $d_S^{crit}$ below which superconductivity vanishes amounts to $d_S^{crit}$ ~ 42 nm for Fe/Cu/Pb, $d_S^{crit}$ ~ 23 nm for Set 1 and $d_S^{crit}$ ~ 12 nm for Set 2. The DSP value for iron has been found by Soulen *et al.* [12] to be 42 %. Our data on the point-contact spectroscopy with Nb tip enable to estimate DSP as $P$ = 30 % for Set 1 and as $P$ = 70 % for Set 2.

The main parameters of the theory by Fominov *et al.* [25] are the transparency parameter $\gamma_b$ (Eq. 4) and the exchange splitting of conduction band of the ferromagnet $h$. To estimate $\gamma_b$ from Fig. 3 (a) we use the values $\rho_s$, $\xi_s$, $\rho_f$ presented above and $\xi_f$ = 14 nm which we roughly estimate basing on our data. From Eq. (3) we obtain $\gamma$ = 0.034 for both sets of samples. Finally, from Eq. (2) we get $\gamma_b$ = 0.15 for Set 1 and $\gamma_b$ = 0.35 for Set 2. With these parameters we have calculated theoretical curves $T_c(d_{Pb})$ in Fig. 3(a). As can be seen there the theory and experiment agree reasonable well. From Eq. (3) one can conclude that the resistance of the S/F boundary $R_b$ for Set 2 is larger by a factor of 2 than the one for Set 1. The question arises if the difference between two sample sets could be due to a more oxidized F/S interface in the sample from the Set 2. As it has been established from the EDXS line scan intensity profiles for the cross-section there are no chemical differences in Heusler/Cu and Cu/Pb interfaces, evidencing the absence of the oxide layer for both types of structures. Hence the additional oxidation or any another chemical modification of the interfaces cannot be the reason for the reduction of the transmission coefficient for electrons moving through the Heusler/Cu/Pb interfaces.

There are two points of view concerning the processes taking place at the S/F interface with high DSP value of the ferromagnetic layer. The first one [26] claims that if a ferromagnet is in the half-metallic regime with only one occupied spin-band, the Cooper pairs with zero spin projection reaching the interface should cease at the interface because one electron of the pair penetrates through the interface into the half-metal while another one with the opposite spin is normally reflected back to the superconductor. Based on our results we conclude that $\gamma_b$ increases with increasing DSP, i. e. the S/F interface becomes less transparent. This means that the Cooper pairs are mostly reflected from the S/F interface without being destroyed [see sketch in Fig. 3 (b)].

The starting point for the second approach is that the theory by Mironov and Buzdin [26] does not account for the possible relative shift between the spin-majority energy band in the half-metal and the electron bands in the superconductor, which may renormalize the probabilities of the electron transmission through the S/F interface. Such kind of effects have been analyzed by Takahashi *et al.* [27]. Following them, we assume a nearly full polarization of conduction band of the Heusler layer in our samples. In that case the resistance of the ferromagnet for the electrons in the Cooper pair is different for spin-up and spin-down electrons. The larger the exchange splitting of the conduction band of ferromagnet is, the larger resistance of the S/F interface is expected. Hence, for larger $h$ values the larger values of $\gamma_b$ are expected. This spin imbalance plays a key role in the processes taking place at the interface. The generalization of the theory by Fominov et al. [25] for strong ferromagnets requires taking into account the fact that the penetration of electrons through the S/F interface occurs with different probabilities for the electrons with different spin projections as it was assumed in the boundary conditions by Eschrig *et al.* [28]. The development of the theory in this direction is appealing for a better, quantitative determination of DSP based on the measurements of the superconducting critical temperature in S/F bilayers.

In summary, we have demonstrated that the dependence of the superconducting critical temperature $T_c$ of the S/F bilayer Pb/Co$_2$Cr$_{1-x}$Fe$_x$Al on the thickness of the S - layer is sensitive to the degree of the spin polarization of the F-layer. Our findings may pave the way to establishing a new advantageous method of determination of DSP in ferromagnetic films beyond already known techniques.

Acknowledgements

We gratefully acknowledge S. V. Mironov and Ya. V. Fominov for stimulating and fruitful discussions. This work was financially supported by the Deutsche Forschungsgemeinschaft through project No. LE 3270/1-2. It was also partially supported by Russian Foundation for Basic Research through project No. 17-02-00229-a and the Program of the Russian Academy of Sciences.